\documentclass[%
 reprint,
superscriptaddress,
prb,
]{revtex4-1}

\usepackage{xcolor}
\usepackage{lineno}
\usepackage{graphicx}
\usepackage{hyperref}
\usepackage{amsmath}
\usepackage{color}
\usepackage{graphics}
\usepackage{adjustbox}
\definecolor{red}{rgb}{0.8, 0.0, 0.0}
\definecolor{blue}{rgb}{0.06, 0.2, 0.65}
\definecolor{green}{rgb}{0,0.6,0}

\begin{document}

\begin{figure}[t!]
    \includegraphics[width=0.80\textwidth]{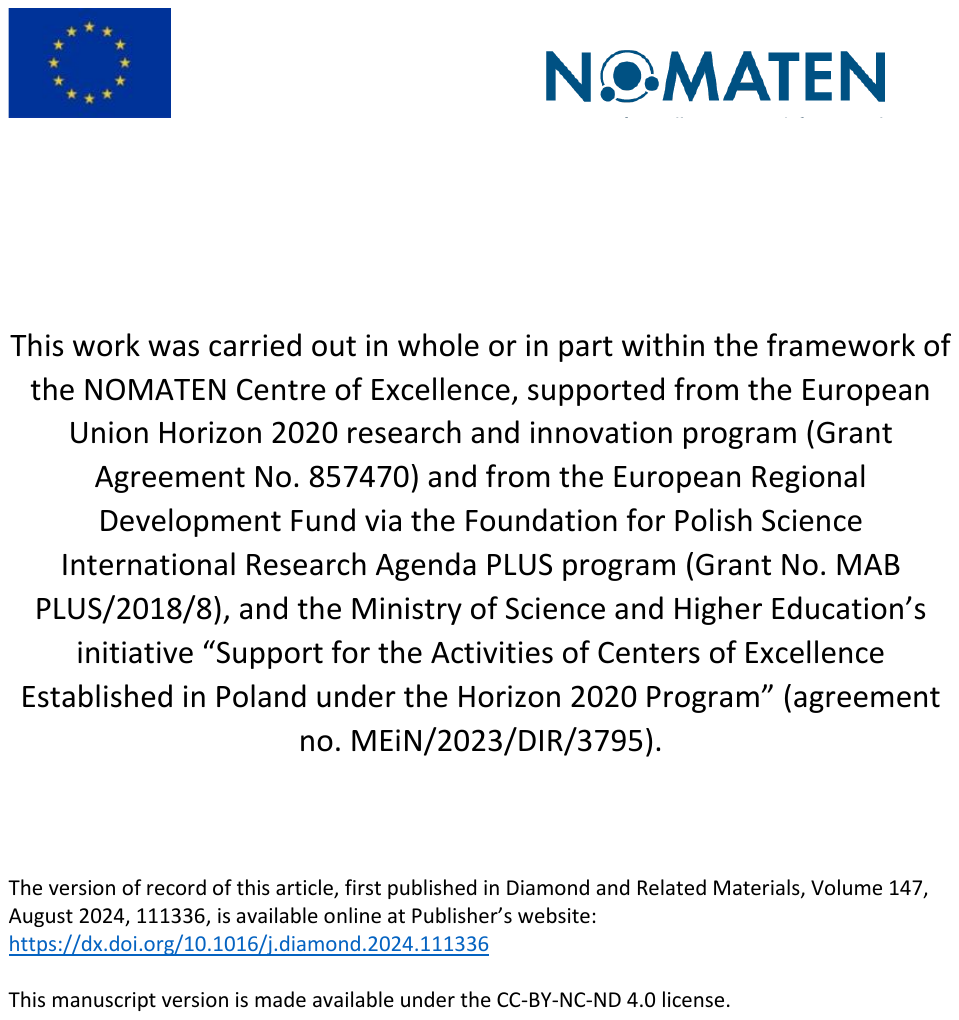}
\end{figure}
\newpage 

\title{CH$_4$ and CO$_2$ Adsorption Mechanisms on Monolayer Graphenylene and their Effects on Optical and Electronic Properties}

\author{A. Aligayev}
\affiliation{Science Island Branch of Graduate School, University of Science and Technology of China, Hefei, 230026, China.}
\affiliation{%
NOMATEN Centre of Excellence, National Center for Nuclear Research, 
05-400 Swierk/Otwock, Poland
 }%

\author{F. J. Dominguez-Gutierrez}
\affiliation{%
NOMATEN Centre of Excellence, National Center for Nuclear Research, 
05-400 Swierk/Otwock, Poland
 }%
\affiliation{Institute for Advanced Computational Science
Stony Brook University Stony Brook, NY 11749, USA}

\author{M. Chourashiya}
\affiliation{%
NOMATEN Centre of Excellence, National Center for Nuclear Research, 
05-400 Swierk/Otwock, Poland
 }%
\affiliation{Guangdong Technion Israel Institute of Technology, Shantou, 515063, China}
\affiliation{Technion, Israel Institute of Technology, Haifa, 32000, Israel}

\author{S. Papanikolaou}
\affiliation{%
NOMATEN Centre of Excellence, National Center for Nuclear Research, 
05-400 Swierk/Otwock, Poland
 }%
\author{Q. Huang}
\affiliation{University of Science and Technology of China, Hefei, 230026, China.}

\vspace{10pt}

\begin{abstract}

Graphenylene (GPNL) is a two--dimensional
carbon allotrope with a hexagonal lattice structure containing 
periodic pores. The unique arrangement of GPNL offers potential 
applications in electronics, optoelectronics, energy storage, and 
gas separation. Specifically, its advantageous 
electronic and optical properties, make it a promising candidate for 
hydrogen production and advanced electronic devices. 
In this study, we employ 
a computational chemistry--based modeling approach to investigate
the adsorption mechanisms of CH$_4$ and CO$_2$ on monolayer
GPNL, with a specific focus on their effects on optical
adsorption and electrical transport properties at room
temperature.
To simulate the adsorption dynamics as closely as possible to 
experimental conditions, we utilize the self--consistent charge 
tight--binding density functional theory (SCC--DFTB). 
Through semi--classical molecular dynamics (MD) simulations,
we observe the formation of H$_2$ molecules from the
dissociation of CH$_4$ and the formation of CO+O species 
from carbon dioxide molecules. 
This provides insights into the adsorption and dispersion
mechanisms of CH$_4$ and CO$_2$ on GPNL.
Furthermore, we explore the impact of molecular 
adsorption on optical absorption properties. 
Our results demonstrate that CH$_4$ and CH$_2$ affects drastically
the optical adsorption of GPNL, while CO$_2$ does not significantly 
affect the optical properties of the two--dimensional material. 
To analyze 
electron transport, we employ the open--boundary non--equilibrium 
Green's function method. By studying the conductivity of GPNL and 
graphene under voltage bias up to 300 mV, 
we gain valuable insights into the electrical transport 
properties of GPNL under optical absorption conditions.
The findings from our computational modeling approach
might contribute to 
a deeper understanding of the potential applications of GPNL in 
hydrogen production and advanced electronic devices.
\end{abstract}

%
\vspace{2pc}
%
%
\maketitle
%
%

\section{Introduction}
\label{sec:intro}


Sensors detect gases through the physical 
adsorption of gas molecules onto a surface. 
These sensors use a gas--sensitive material, such as 
carbon--based ones, that change their properties 
upon gas adsorption, offering high sensitivity and 
selectivity \cite{JANA2022108543}. 
The high sensitivity of graphene to the local environment
has shown to be highly advantageous in sensing applications,
where ultralow concentrations of adsorbed molecules induce
a significant response to the electronic properties of
graphene 
\cite{balandin2011thermal,Wang_2022,PhysRevB.77.125416,C6CP07654H,SHABAN20194510,ALIGAYEV2022355}. 
Additionally, carbon--based materials can be tailored by
varying their surface chemistry, porosity, and morphology.
The hybridization of carbon atoms into sp--, sp$^2$--, and
sp$^3$--orbitals can create diverse carbon allotropes
exhibiting distinct dimensionalities \cite{shanmugam2022review,gao2023tunable}. 
Ongoing research focuses on improving the performance and 
reliability of gas adsorption sensors and exploring 
new materials and sensing mechanisms for enhanced gas sensing 
capabilities 
\cite{hirsch2010era, karfunkel1992new, baughman1987structure, li2009superhard, sheng2011t,C5CP04422G}.

Graphenylene (GPNL), an intriguing carbon allotrope sharing
the same point group (D$_{6h}$) as graphene, is composed of 
sp$^2$--hybridized carbon atoms arranged in hexatomic and
tetratomic rings \cite{balaban1968chemical, balaban1994diamond,C2TC00006G, D0CP04188B}. 
Qi-Shi Du et al. successfully synthesized layers of graphenylene, also referred
to as biphenylene--carbon (BPC), by dehydrating and polymerizing 1,3,
\newline
5--trihydroxybenzene \cite{du2017new,zhang2019art}. 
The process involved the removal of three water molecules
from a 1,3,5--trihydroxybenzene molecule using dehydrant aluminum oxide, leading to the
amalgamation of bare 6C rings (benzynes) and the formation of a small segment of
the 2D carbon crystal. Furthermore, polymerization could also take place through
intermolecular dehydration between 1,3,5-trihydroxybenzene molecules, 
where the fragments of the 2D carbon crystal grew rapidly. The experimental
construction of GPNL involved utilizing planar 4-carbon rings and 6-carbon rings
with sp$^2$ electron configuration, experiencing slight distortions that ultimately
resulted in the formation of a large planar conjugated $\pi$-system \cite{brunetto2012nonzero}.
It possesses a hexagonal lattice structure with periodic 
pores, offering a high surface area and pore volume. 
These characteristics make GPNL a promising 
material for gas adsorption and separation applications.
The unique topology of GPNL allows for selective 
adsorption of specific gas molecules, making it a potential 
candidate for highly efficient and selective gas separation
and  storage processes 
\cite{KOCHAEV2021109999, Martins2022,D0RA04286B}. 
The identification of hollow adsorption 
sites in GPNL is of great interest, as these sites hold 
significant potential for various applications, including gas 
separation \cite{xu2017inorganic, zhu2016theoretical, 
rezaee2020graphenylene, motallebipour2021graphenylene}.

The growing importance of air quality and 
safety has created a demand for advanced 
gas sensors. Porous carbon--based materials 
have emerged as promising candidates for 
such sensors due to their comparable 
electronic mobility and mechanical 
properties to graphene. Additionally, 
these materials offer the added advantage 
of enabling the dispersion of single atoms 
within acetylenic pores.
Building upon the research progress in graphene, 
investigations into post--graphene 2D carbon--based materials 
have swiftly demonstrated diverse electronic devices
and emerging charge transport phenomena. 
However, despite the growing understanding of electronic 
transport in individual 2D materials, practical 
wafer--scale implementation faces significant challenges 
\cite{Sangwan,D1CP01890F}. 
Therefore, the development of reliable techniques for 
wafer--scale growth, ensuring uniformity and predictable 
thickness poses considerable hurdles from a materials
science perspective. 
Thus, carbon materials exhibit versatile bonding abilities,
ranging from sp$^1$ to sp$^3$ hybridization, and encompass
a variety of allotropes, including fullerene, graphite,
diamond, graphene, carbon nanotubes, and fibers 
\cite{CHUNG2023215066,WandOganov,Keith}. 
Porous carbons can be obtained through the carbonization
of natural or synthetic precursors, followed by activation,
enabling tunability of pore sizes across a wide range, 
from micropores ($<$ 2 nm) to mesopores (2-50 nm) and
macropores ($>$ 50 nm). 
Diverse synthesis strategies, such as template methods,
etching of metal carbides, and sol--gel processing, 
have been explored to create porous carbon materials
with controlled pore structures at both the 
micropore and mesopore levels \cite{CHUNG2023215066}. 
These porous carbons find applications in crucial fields
such as adsorption, separation, and electrode materials.

GPNL, with its exceptional porous architecture and 
remarkable electronic features hold great promise as a 
material for the development of high--performance molecular
gas sensors. In order to save financial resources and avoid 
exhaustive experimental trials, detailed atomistic simulations 
are essential to complement practical exploration.
In this study, we employ the Self--Consistent--Charge Density--
Functional Tight--Binding (SCC--DFTB) method \cite{DFTBplus, Qiang} 
to investigate the potential applications of GPNL in the 
detection of important molecular gases such as CH$_4$ and CO$_2$ 
and their species, which have significant environmental
implications \cite{Santos_2021, ARUNRAGSA2020107790}. 
Here we investigate the hydrogen production through methane 
dissociation and CO$_2$ reduction mechanisms by emitting 
CH$_4$, and CO$_2$ molecules with impact energies close to
their dissociation energies and studying their interactions with 
GPNL. Additionally, we also compare our results with the findings 
obtained for graphene. Further computational research that 
closely emulates dynamic mechanisms observed in experiments is 
necessary to fully explore the potential of graphenylene, 
including its effects on optical absorption, electron transport 
performance, and enhancement of material sensitivity. Our primary 
objective is to contribute to the characterization of 
GPNL as a promising material for future research and the 
development of materials for ultrafast gas sensors and gas separation 
applications.
\section{Computational Methods}
\label{sec:methods}

The SCC--DFTB method is a computational approach that approximates 
traditional Density Functional Theory (DFT) by considering 
valence electron interactions in MD simulations. 
It serves as a valuable tool for accurately predicting structures 
and thermodynamic properties prior to synthesis, providing 
insights into the gas adsorption properties of 2D carbon--based 
materials and their potential applications in various gas 
adsorption environments.
The SCC--DFTB method involves solving Kohn--Sham equations to 
obtain total valence electronic densities and energies for each 
atom utilizing a Hamiltonian functional based on a two-center 
approximation and optimized pseudo--atomic orbitals as the basis 
functions \cite{DFTBplus, Qiang}. Slater--Koster parameter files 
are utilized to provide tabulated Hamiltonian matrix elements, 
overlap integrals, and repulsive splines fitted to DFT 
dissociation curves. 
These parameters describe the overlap and hopping integrals 
between pairs of atoms in the tight--binding Hamiltonian. 
The optimal set of Slater-–Koster parameters have two main 
requirements: a good reproduction of the structure of the 
relevant electronic bands, and faithful representation of the 
orbital contribution along such bands.
Therefore, in the scope of this approach the total energy
of the system is expressed as
\begin{equation}
    E^{\rm DFTB} = E_{\rm band}+E_{\rm rep}+E_{\rm SCC},
\end{equation}
with the band structure energy, $E_{\rm band}$, defined
from the summation of the orbital energies $\epsilon_i$
over all occupied orbitals $\Psi_i$; 
the repulsive energy $E_{\rm rep}$ for the core--core
interactions related to the exchange--correlation energy 
and other contributions in the form of a 
set of distance--dependent pairwise terms;
and an SCC contribution, $E_{\rm SCC}$, as the contributions
given by charge--charge interactions in the system.
Therefore, the electronic energy is calculated by summing 
the occupied Kohn-Sham (KS) single--particle energies and the 
contributions from repulsive energies between diatomic atoms. 
To account for self--consistent charge (SCC) effects during 
the dynamics, an iterative procedure is used. 


\subsection{Structures and binding energies}

\begin{figure}[b!]
    \centering
    \includegraphics[width=0.5\textwidth]{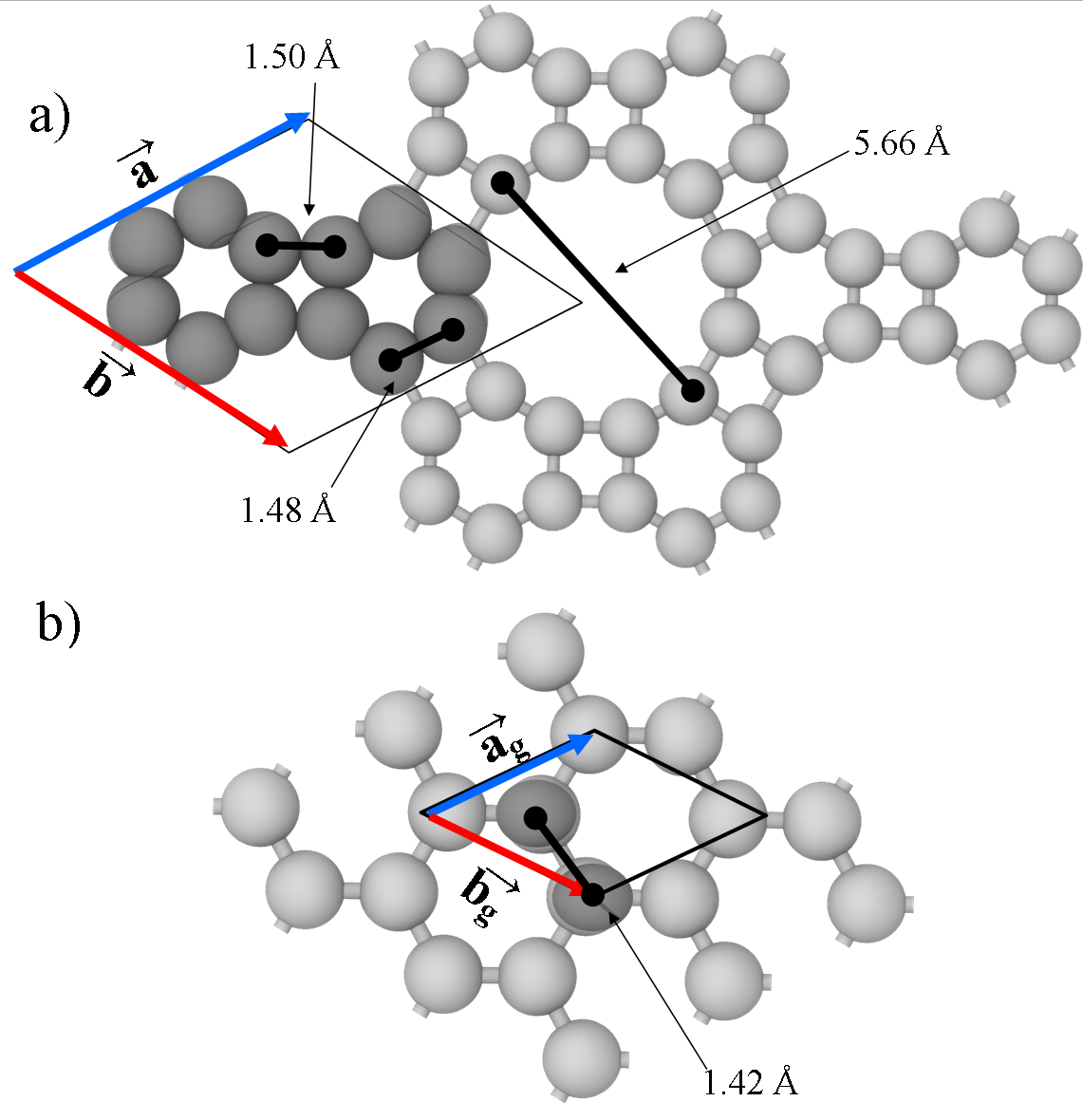}
    \caption{(Color online). Optimized structures of graphenylene (a) and graphene (b) were obtained using the SCC--DFTB method. The calculated bond lengths and lattice parameters are in good agreement with the reported DFT data 
    \cite{fabris2018theoretical}.}
    \label{fig:structureGPNL}
\end{figure}

GPNL is a two--dimensional carbon allotrope 
that possesses a hexagonal lattice structure with periodic
pores and its structure as reported 
by Balaban et al. \cite{balaban1968chemical}  and Martins et al. 
\cite{Martins2022}, consists of three types of symmetrically 
distributed rings: dodecagon (C$_{12}$), hexagon (C$_6$), and
square (C$_4$), which forms a tiling 
of the Euclidean plane. The unit cell of GPNL, determined by 
Fabris et al. using DFT\cite{fabris2018theoretical}, belongs
to the $P6/mmm$ space group and contains a single
irreducible atom, which is considered in our SCC--DFTB 
calculations.
In our study, we performed optimization of the GPNL unit 
cell, resulting in lattice parameters 
$\vec{ \text{a}} = \vec{b} = 
6.735$ \AA{} and bond lengths of 1.50 Å for the square ring and 1.48 
Å for the hexagon ring identifying seven points of 
high symmetry \cite{martins2021new} which is 
in good agreement with experimental measurements for 1.42--1.46 \AA{} for the 6--C rings and 1.50-1.52 \AA{} for the two bonds 
joining the 6--C rings \cite{du2017new},  as shown in Fig. \ref{fig:structureGPNL}a). 
The central nanopore (dodecagon ring) in the unit cell has a diameter 
of 5.66 Å, in good agreement with DFT data  
\cite{fabris2018theoretical,C2TC00006G} 
and experimental measure of 5.8 \AA{} \cite{du2017new}. 
To compare the adsorption capabilities of GPNL with graphene, 
we also optimized the unit cell of graphene using a well--known 
lattice parameters and bond lengths, as depicted in Fig. 
\ref{fig:structureGPNL}b).

The interaction potentials between H$_2$, CO$_2$, and CH$_4$ 
molecules with graphene and GPNL are investigated using the 
DFTB method. To avoid interactions with periodic replicas, the unit 
cell of the optimized GPNL structure is replicated by 
3$\times$4$\times$1 and the unit cell of graphene is replicated by 
5$\times$5$\times$1 along the $x$ and $y$ directions. 
The larger cells are initially optimized using SCC-DFTB. 
Adiabatic calculations follow to determine potential energy curves. 
These curves depict molecule interactions with fully relaxed periodic 
sheets at different distances and adsorbate sites, incorporating 
dispersion corrections through van der Waals interactions.
\cite{10.1063/1.1329889}.
Thus, the total energies, $E(z)$, of the molecule--2D material 
system with a separation $z$ between the adsorbate 
sites and the center of mass of the molecules
are varied above the surface in a range of 0.5 to 7 \AA{}, 
which defines the computation of the adsorption potential 
as a function of the distance separation.
The total energy is then computed as:
\begin{equation}
    E(z) = E_{\rm Tot} - \left( E_{\rm 2D material} + E_{\rm Molecule}
    \right),
\end{equation}
where $E_{\rm Surface}$ is the total energy of the 
2D material; $E_{\rm Molecule}$ is the total energy of the 
isolated molecule: H$_2$, CH$_4$, 
and CO$_2$; and $E_{\rm Tot}$ is the energy of the 
interacting system at every $z$--distance. 
Thus, the binding energy is defined as 
$E_b = E(z_{\rm min})$ with $z_{\rm min}$ as the 
equilibrium molecule--surface distance. 
Total energy calculations are performed for the 
molecule--2D material system, varying the distance between 
the surface and the center of mass of the molecules 
along the $z$--axis. 
We consider 3 different adsorption sites for graphene and 
5 sites for GPNL based on unit cells of 
the materials.
The molecular symmetry planes are consider 
with respect to the surface plane in the calculations. 
The repulsive potential is cut off at
a distance below the second nearest--neighbor interaction
region for numerical stability. 

\subsection{Semi--Classical Molecular Dynamics Simulations}

We conducted semi--classical molecular dynamics simulations
to investigate the adsorption dynamics of H$_2$, CH$_4$,
and CO$_2$ molecules on graphene and GPNL. 
For graphene, a $5\times5\times1$ supercell 
was used, while for GPNL, a $3\times3\times1$ supercell was 
employed. The surfaces were optimized and equilibrated to a 
temperature of 300 K using a Nose--Hoover thermostat.
To simulate the adsorption dynamics, we defined a target area of 1 
nm$^2$ on the surface, and molecules were randomly distributed on it 
using the velocity Verlet algorithm. 
The impact energy of the molecules was 8 eV, 
and 650 independent trajectories were generated for each molecule. 
A time step of 0.25 fs was used, and the molecules were emitted 
vertically with random orientations at an initial distance of 
0.6 nm above the surface. The simulations were 
performed for a duration of 350 fs,  
This timeframe was meticulously chosen to ensure convergence
in our MD simulations. It provided ample time for the molecules
to travel away from the carbon sheets while also guaranteeing
that the attached molecules remained bonded to the carbon atoms, 
preventing any detachment from the sheets.
We have previously employed this approach to study hydrogenation 
mechanisms of fullerene cages \cite{DOMINGUEZGUTIERREZ2018189}, 
electronic properties 
of borophene \cite{C7TC00976C}, and dynamic physisorption pathways of 
molecules on alumina surfaces \cite{aluminadftb}, demonstrating 
excellent agreement with first principles DFT calculations.

\subsection{Optical absorbance and electron transport calculations}
%

The optical absorption is investigated within the DFTB framework as 
an electronic dynamic process in response to an external electric
field \cite{C8CP04625E,B926051J}. 
The conventional adiabatic approximation gives the time evolution 
of the electron density matrix by time integration of the 
Liouville--von Newmann equation expressed as
\begin{equation}
    i \hbar \frac{\partial \hat \rho}{\partial t} = 
    S^{-1}\hat H \hat \rho - \hat \rho S^{-1},
\end{equation}
where $\hat \rho$ is the single electron density matrix, 
$\hat S$ is the overlap matrix, and $\hat H$ is the system Hamiltonian 
that includes the external electric field as 
$\hat H = \hat H_0 + E_0 \delta (t-t_0) \hat e$ with 
$E_0$, the magnitude of the electric field, 
and $\hat e$, its direction. 
Under the framework of linear response, the absorbance $I(\omega)$ is
calculated as the imaginary part of the Fourier transform of the
induced dipole moment caused by an external field. 
In this study, the external field strength was set to 
$E_0 = 0.001$ V/\AA{}. The induced dipole moment was evaluated over
a $200$ fs time period using a time step of $\Delta t$ = 0.01 fs. 
The Fourier transform was performed with an exponential damping
function (using a 5 fs damping constant) to eliminate noise.

\begin{figure}[t!]
    \centering
    \includegraphics[width=0.45\textwidth]{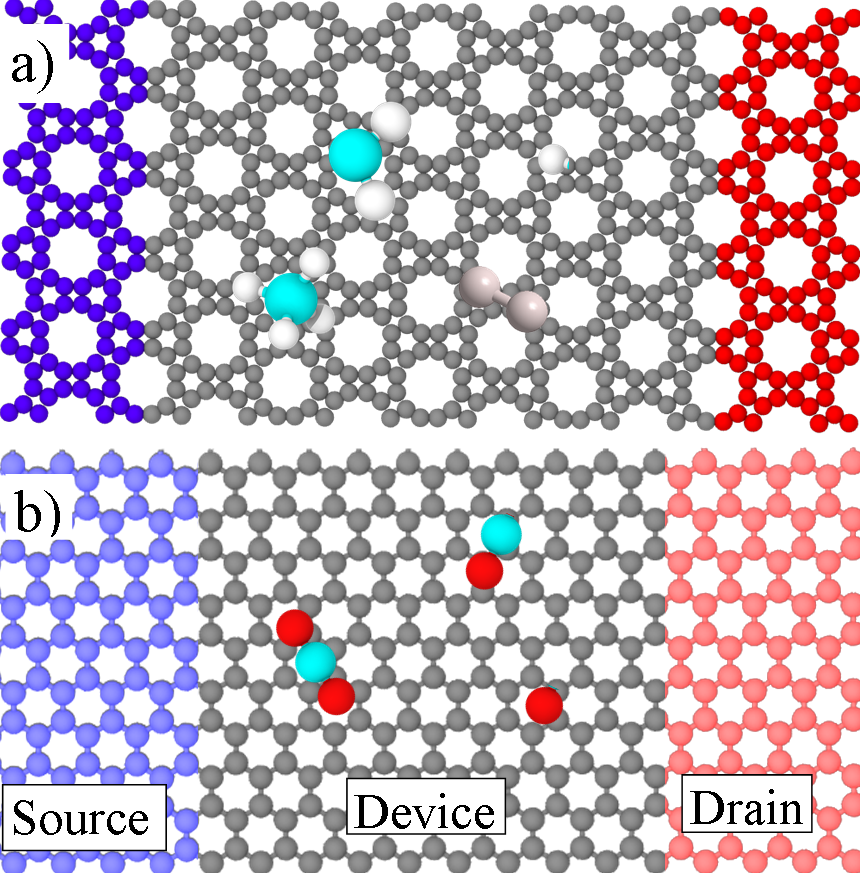}
    \caption{Optimized structures employed in the electron 
    transport calculations. In (a), the structure is shown for 
    graphenylene with CH$_4$, CH$_2$, and H$_2$ molecules and 
    a hydrogen atom attached, while in (b), graphene 
    sheet is depicted with CO$_2$ and CO molecules 
    with an oxygen atom. 
    The blue and red regions represent the two 
    principal layers. The middle region corresponds to the device under
    investigation and the size of the molecules is increased for better visualization.}
    \label{fig:transport_}
\end{figure}

 The Non--Equilibrium Green's Functions formalism (NEGF) 
 is a robust  theoretical framework commonly used for 
 modeling electron transport in  nano--scale devices and
 is implemented in the DFTB code  \cite{Pecchia_2008}. 
In Fig. \ref{fig:transport_}, we provide a detailed illustration
of the geometric configuration of the graphene and GPNL 
structures, highlighting the specific regions involved in the 
electron transport calculations. To ensure accurate and reliable
results, several steps are followed: 1) The structures are carefully
divided into distinct sections, including the principal layers, 
two electrode contacts (drain and source), and the device region. 
This partitioning enables a systematic analysis of electron 
transport within the designated "scattering region."; 2) 
The drain section, represented by red spheres, corresponds to 
the region where electrons exit the device, while the source 
section, depicted by blue spheres, represents the region where 
electrons enter the device; 3) To simulate realistic conditions and 
investigate the impact of specific molecules on electron transport,
CO$_2$ molecules are introduced into the graphene device section,
and CH$_4$ molecules are added to the GPNL device section.
This allows us to study the interaction between the adsorbates
and the carbon-based materials and observe their influence on
the electron transport properties; and 4) Before performing the 
electron transport calculations, the entire system undergoes an 
optimization process. This optimization involves adjusting the 
positions and orientations of the atoms to find the most 
energetically favorable configuration for the combined 
graphene/GPNL--adsorbate system.

\section{Results}
\label{sec:results}

\begin{figure}[b!]
    \centering
    \includegraphics[width=0.48\textwidth]{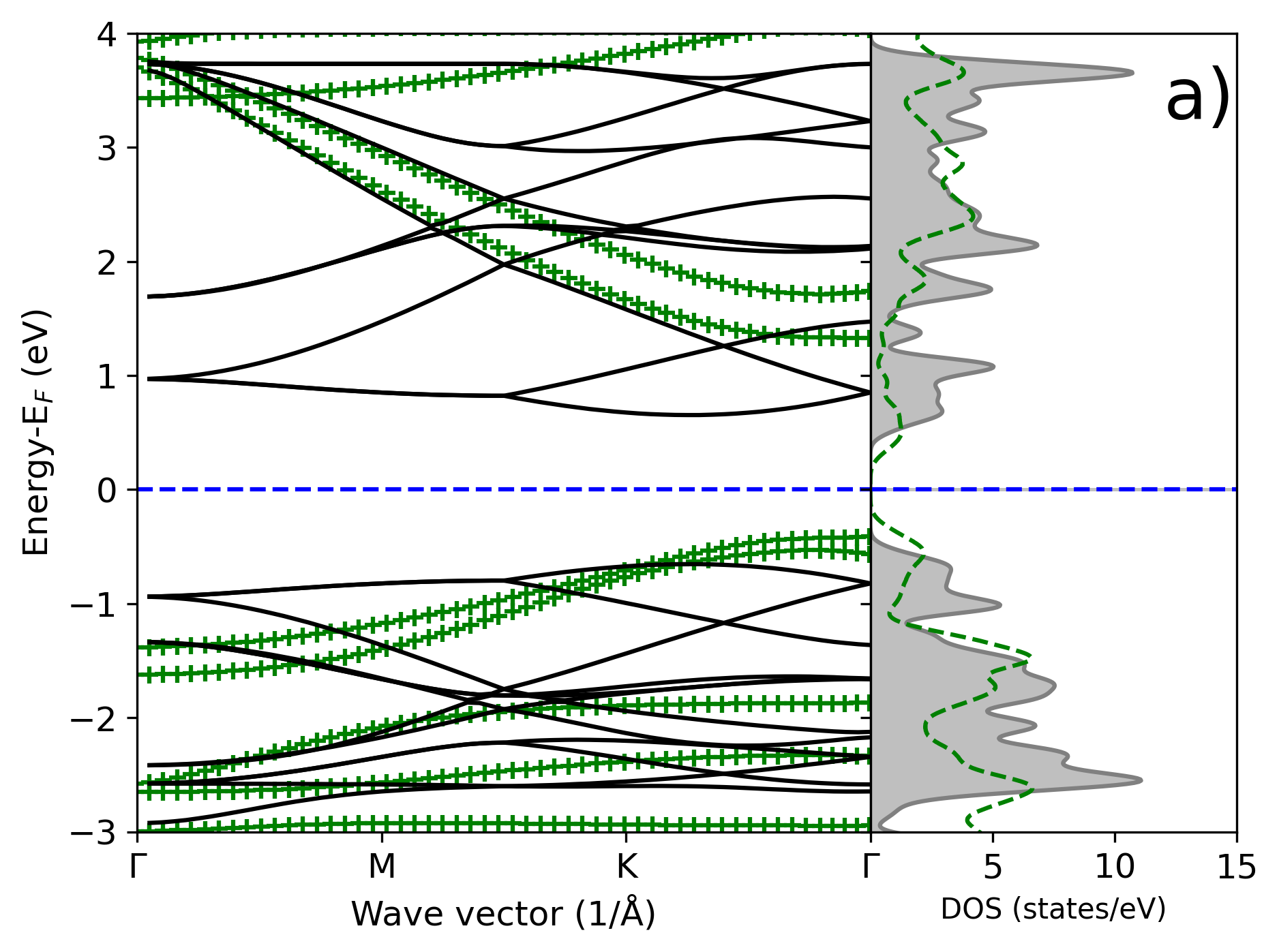}
    \includegraphics[width=0.48\textwidth]{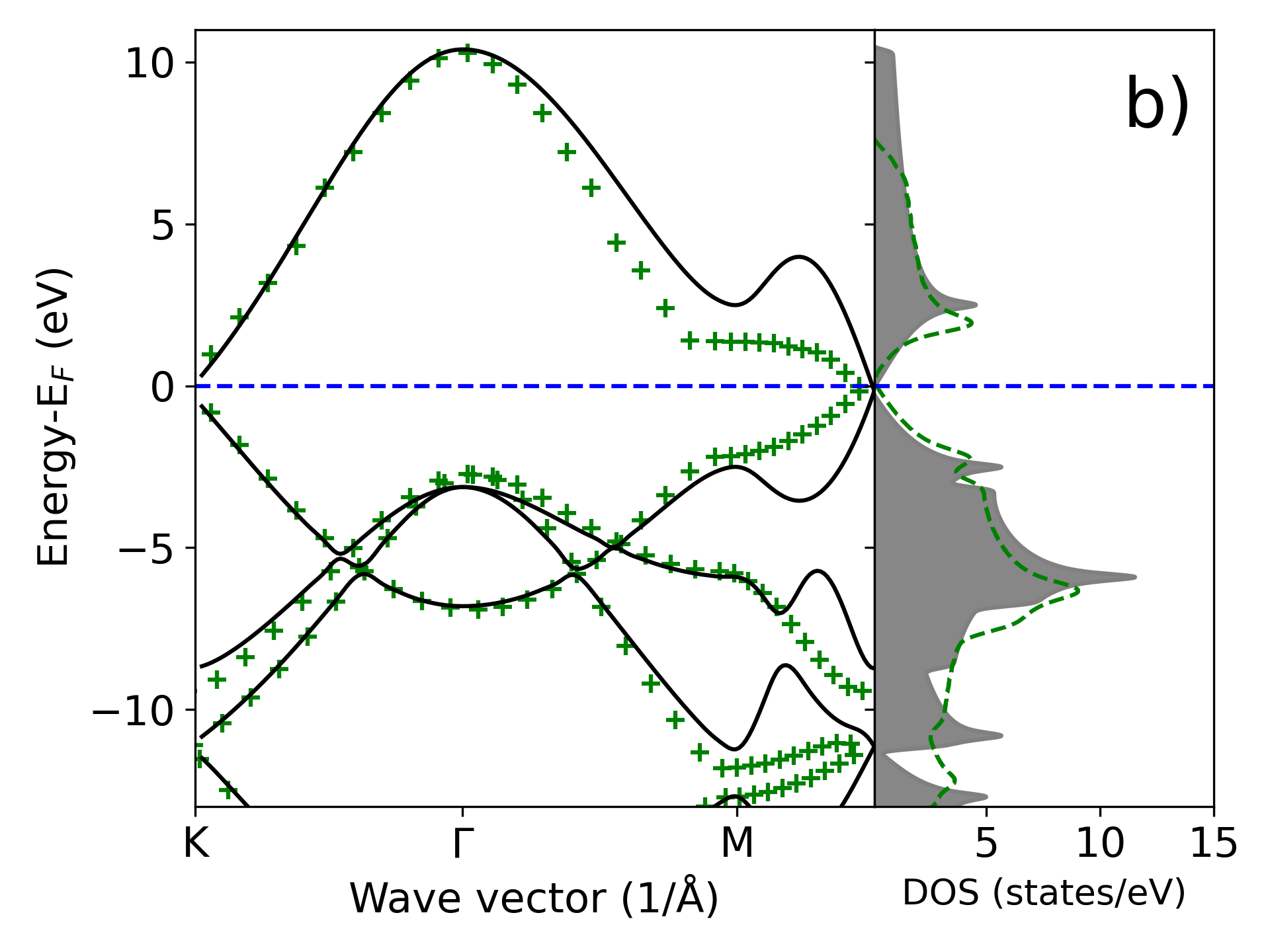}
    \caption{Band structure and density of states for GPNL in a) 
    and for graphene in b). Similarities are observed due 
    to the hexagonal arrangement in the unit cell of the 
    materialsi. DFT calculations are represented by dashed lines 
    for the density of states (DOS) and marked with cross-points 
    for the band structure.}
    \label{fig:DOS}
\end{figure}


GPNL exhibits lower electric conductivity compared to
graphene due to its distinct physical properties, particularly
its formation energies and band gaps. The 
influence of pore size and quantity on these properties is 
depicted in Fig. 
\ref{fig:DOS}. Our DFTB calculations reveal that GPNL 
possesses a bandgap of approximately 0.96 eV from 
the DOS calculations, while graphene lacks a bandgap altogether.
The porosity of GPNL has the potential to significantly modify its 
electronic characteristics and catalytic performance by increasing
its surface area, in good agreement with reported 
results by G. Brunetto et al \cite{brunetto2012nonzero}. The 
electronic band structures, illustrated in Fig. \ref{fig:DOS},
reveal that the valence and conductance
bands for carbon atoms are located at the $\Gamma$ point.
The band gap of GPNL is structure--dependent and can range from
zero to a few electron volts. Theoretical studies have predicted
band gaps for GPNL ranging from 0 eV to approximately 3.3 eV,
depending on the specific structure and calculation 
method employed \cite{balandin2011thermal,brunetto2012nonzero}.
It should be noted that the semi--local functionals tend to 
underestimate the band gaps of GPNL structures. 
It is also noticed the selected path show the characteristic
gaps at the M and $\Gamma$ point reported by DFT 
calculations \cite{C2TC00006G} and our results reported in the SM. 
The slight discrepancy in negative energies can be attributed to 
variations in the SK parameters applied in our calculations and the 
pseudo potentials used in our DFT calculations. Nevertheless, this 
fair agreement still provides validation for our results.

To validate our findings, density functional theory (DFT) 
calculations were conducted using the PBE exchange--correlation 
functional. The calculations were carried out under periodic 
boundary conditions, and the Brillouin zone integration was 
performed with the $\Gamma$ point considered. Kohn--Sham orbitals 
were employed as plane waves up to an energy cutoff of 90 Ry to 
ensure convergence in the structural properties of the systems. 
The Quantum-ESPRESSO ab--initio package with relativistic-
corrected pseudo-potentials was utilized for computing the 
density of states, system energies, and band structures.
The total electronic density of states (DOS) for GPNL
reveals significant overlaps between the C--2s and C--2p curves, 
indicating the presence of strong sp$^3$ hybridized covalent
bonding states.

GPNL consists of interconnected benzene 
rings arranged in a hexagonal lattice, similar to graphene.
The slight discrepancy in negative energies for the GPNL sheet
can be attributed to variations in the SK parameters applied in our 
calculations and the pseudo potentials used in our DFT calculations. 
Nevertheless, this fair agreement still provides validation for our 
results.
This supports the application of SCC--DFTB in studying gas
separation processes, which is crucial for the production
and utilization of clean fuels. 
Different paths for the GPNL sheet was considered 
and shown in the supplementary material (SM) of this work.


Figure \ref{fig:PECs} illustrates the binding energies as a
function of separation distance for graphene (a) and GPNL (b), considering different adsorbate sites as
labeled in the inset figure. We performed adsorption 
calculations for isolated H$_2$, CH$_4$, and CO$_2$ molecules
on both systems.
To ensure accurate calculations, we included a 50 \AA{}
vacuum section above the sample to minimize boundary effects.
Periodic boundary conditions were applied in the $x$--$y$
directions to simulate a semi-infinite surface. 
For the $k$--point sampling, we employed a $4\times4\times1$
Monkhorst--Pack set throughout all calculations.

\begin{figure}[b!]
    \centering
\includegraphics[width=0.48\textwidth]{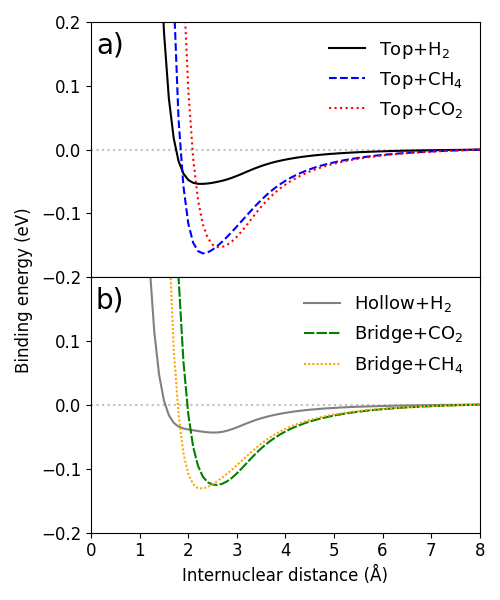}
    \caption{Binding energies are presented as a function of 
    separation distance for both graphene (a) and graphenelyne (b), 
    showcasing the lowest energy states and the impact of distinct 
    adsorbate sites on the binding of H$_2$, CO$_2$, and CH$_4$ 
    molecules. The analysis reveals a stronger tendency for 
    adsorbate molecules to form bonds with the top carbon atoms in 
    graphene, while graphenelyne displays diverse adsorbate sites 
    due to its sp hybridization. The remaining potential energy 
    curves (PECs) can be found in the Supplementary Material 
    accompanying this manuscript.}
    \label{fig:PECs}
\end{figure}

Fig. \ref{fig:PECs} present results for the physisorption 
pathways for H$_2$, CH$_4$, and CO$_2$ molecules on both
graphene and GPNL sheets. In the case of graphene, we 
observe that the bond length for molecular hydrogen and 
methane molecules is approximately 2.2 \AA{}, with binding 
energies of -0.052 eV (in good agreement with Lee et al. \cite{10.1063/5.0116092}) and -0.165 eV, respectively. 
Meanwhile, the carbon dioxide molecule exhibits a bond 
length of 2.6 \AA{} and a binding energy of -0.151 eV.
For GPNL, we find that H$_2$ and CH$_4$ molecules have bond 
lengths of 2.25 \AA{}, with binding energies of -0.043 eV 
and -0.132 eV, respectively. CO$_2$ on GPNL has a bond 
length of 2.62 \AA{}, accompanied by a binding energy of 
-0.125 eV. 
In the SM, we present the physisorption pathways of all 
the adsorbate site of both carbon sheets, showing that 
adsorbate sites associated with
the hexagonal and square holes exhibit a lower likelihood for 
molecule adsorption; specially methane molecules are more 
likely to bond to the top of the carbon atoms, with a binding
energy of -0.2 $\pm$ 0.05 eV showing excellent conditions
for gas separation for this energy barrier 
\cite{C2TC00006G,D0RA04286B}.

These results serve as the basis for configuring
the initial conditions in our 
MD simulations. We ensure an initial distance greater
than 8 \AA{} to prevent interactions between the 
molecules and the carbon sheet at the outset of the
simulation.
These findings offer valuable insights into the interactions 
between carbon sheets and these molecules, which are essential 
for understanding the physical processes in our MD simulations.
It is worth noting that dispersion 
corrections
play a significant role in these calculations, particularly due
to the hybridization of the system and the presence of CH$_4$
molecules.
Similar trends are observed for graphene, where the adsorbate site
located at the top of the carbon atoms is more favorable for both 
physisorption and chemisorption mechanisms confirming reported 
results for hydrogen trapping by graphene 
\cite{10.1063/1.1329889}, 
indicating a higher likelihood of attracting molecules.

\subsection{Dynamical adsorbtion of CH$_4$ and CO$_4$ molecules}

After conducting MD simulations at room temperature 
and an impact energy of 8 eV, the emission of hundreds of H$_2$, 
CO$_2$, 
and CH$_4$ molecules is analyzed by calculating the 
probabilities of adsorption, reflection, and transmission, which are 
defined as:
\begin{equation}
    P = 100 \times \frac{N_{x}}{N_{\rm Tot}},
\end{equation}
where $N_{\rm Tot}$ is the total number of MD simulations, while 
$N_x$ is the number of cases for adsorption, reflection, and 
transmission calculated by using the following conditions based 
on the last frame of the MD simulations: 
1) Adsorbed cases: Molecules with a final position between
a sphere centered at the material with a 
radius of 3.5 \AA{} and the direction of the velocity vector
points towards the surface is considered as adsorbed; 
2) Reflection cases: Molecules 
with a final position larger than 3.5 \AA{} and a velocity vector
oriented in the opposite direction
to the surface normal is counted as reflected; 3) 
Transmission cases: Molecules with a final position below the
surface and a distance larger than -3.5 \AA{} are counted as
transmitted. 
The counts for these cases are tabulated in Table \ref{tab:prob}.

\begin{table}[t!]
    \centering
     \resizebox{\columnwidth}{!}{%
    \begin{tabular}{llll| lll}
    \hline
    \multicolumn{4}{c}{Graphene} & \multicolumn{3}{c}{Graphelyne} \\
    \hline
    Probability    &   H$_2$ & CO$_2$ & CH$_4$ & H$_2$ & CO$_2$ & CH$_4$ \\
    \hline
     Transmission  & 0 & 0   & 0    & 5.28   & 1.28  & 49.12 \\
     Adsorption   & 0 & 0   & 16.8 & 0      &  9.66 & 1.60 \\
     Reflection    & 100 & 100 & 83.2 & 94.72  & 89.06 & 49.28 \\
    \hline
    \end{tabular}}
    \caption{probabilities of different cases observed in the MD 
     simulations, including transmission, reflection, and adsorption, 
     for both graphene and GPNL. The results clearly demonstrate 
     that GPNL exhibits superior performance as a material for 
     gas separation compared to graphene.}
    \label{tab:prob}
\end{table}

For graphene, we did not observe dissociation of H$_2$ and CO$_2$ 
molecules among the reflected molecules. 
The molecule initiates its trajectory with a 
kinetic energy (KE) of 8eV. However, it undergoes a gradual 
deceleration as it comes into contact with the charge
cloud of the carbon sheets. Upon colliding with a carbon
sheet at various adsorbate sites, the molecule is reflected, 
promoting its vibrational and rotational movements.
To observe dissociation mechanism, a
higher impact energy than 10 eV would be required.
However, it is worth noting that 
such high--energy collisions could result in the creation of 
vacancies in the graphene sheet by displacing a carbon atom, 
which is not observed in our MD simulations at 
8 eV. On the other hand, some 
of the reflected CH$_4$ molecules underwent dissociation, 
and a few of them were able to attach to the graphene sheet.
In order to bond more 
CH$_2$ molecules to graphene, a lower impact energy would be 
sufficient.
In the case of GPNL, its inherent porosity allowed for a higher 
number of transmitted cases for both H$_2$ and CO$_2$ molecules. We 
observed that some carbon dioxide molecules dissociated into CO+O, 
with 
the CO molecules bonding to the GPNL sheet. This highlights the 
advantage of GPNL's porous structure in facilitating gas 
transmission and reactivity compared to graphene.

\begin{figure}[t!]
    \centering
    \includegraphics[width=0.48\textwidth]{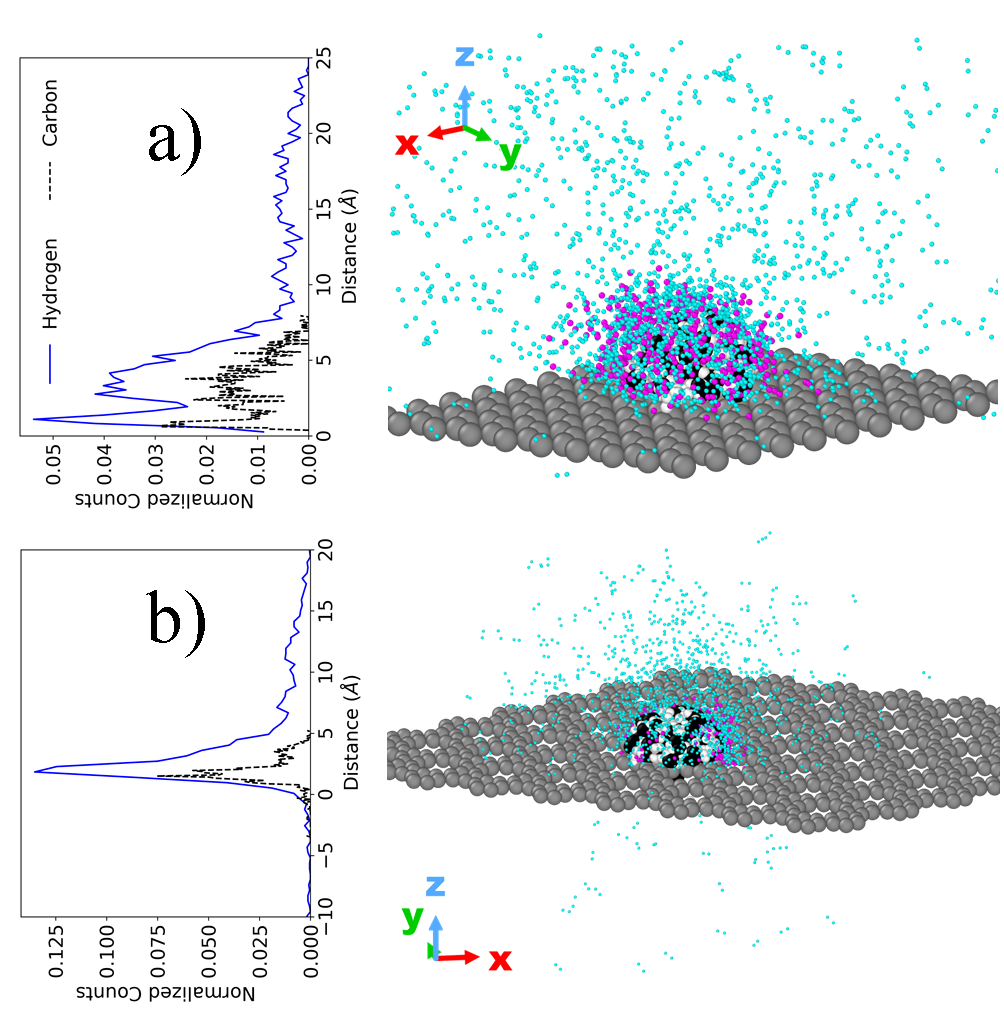}
    \caption{
    The visualization of the final frame in each MD 
    simulation simultaneously enables a thorough analysis of the 
    adsorbed, reflected, and transmitted cases for both Graphene (a) and 
    graphenelyne (b); with their corresponding histogram for the distance counts on the z-axis. The structures consist of carbon atoms depicted as 
    gray spheres. Adsorbed CH$_2$ molecules are represented by black 
    spheres for carbon atoms and white spheres for hydrogen atoms. 
    Reflected molecules are shown as turquoise spheres representing 
    molecular hydrogen, and purple spheres representing carbon atoms.}
    \label{fig:dynamics}
\end{figure}

From our MD simulation results, we have observed that CH$_4$ molecules 
can undergo reflection and dissociation, leading to the formation of 
CH$_2$ and H$_2$ as the main process. Figure \ref{fig:dynamics} 
displays a histogram of the counts of the final positions 
of the C and H atoms for the final frames of the simulations for 
graphene (a) and GPNL 
(b), providing a visual representation of the observed dynamics. 
This plot is used to count for the number of events 
with a probability of transmission, reflection, or absorption.
In the figures, we correlated the visualization 
all the events to the histogram by shown the last frame of all the 
MD simulations in a single image where 
C atoms forming graphene and GPNL are represented as 
gray spheres, adsorbed CH$_2$ molecules are depicted as black 
spheres 
for C atoms and white spheres for hydrogen atoms, while reflected 
molecules are represented by turquoise spheres for molecular 
hydrogen 
and purple spheres for C atoms. 
Within our observations, we find that CH$_2$ 
molecules are more likely to binding to the carbon sheets. In 
alternative scenarios, both CH$_2$ and H$_2$ molecules are seen
to dissociate after collision with the carbon sheets. A notable 
observation is the differential behavior of hydrogen molecules in 
graphene compared to GPNL. 

The GPNL's porous structure facilitates 
the transmission of hydrogen molecules, making it a promising 
candidate for applications such as hydrogen production in the 
context of energy generation. While graphene sheets present 
high scattering of H$_2$ molecules.
We have found that CH$_4$ molecules are unable to transmit through 
graphene, whereas in the case of GPNL, there is a probability of 
approximately 50\% for transmission, accompanied by dissociation 
and the 
production of molecular hydrogen. This behavior can be attributed 
to the 
porous nature of the materials and the impact energy of the 
emitted. 
In addition, we analyze the velocity 
distributions 
of carbon (C), oxygen (O), and hydrogen (H) atoms at the last
frame considering all the MD simulations for each case. 
Notably, molecular hydrogen exhibits the highest velocities, 
while CH$_2$ molecules exhibit the lowest 
velocities after being reflected by the carbon sheets. 
This observation provides an initial indication of the bonding 
behavior of the ejected molecules, which is discussed in more
detail within the SM.

Furthermore, we have noted that CH$_2$ molecules can form bonds 
with GPNL by infiltrating the porous structure and binding to the 
underlying C atoms. This property distinguishes GPNL from 
graphene, as the latter typically requires the presence of
defects to enable molecule transmission. 
Here, the distinctive arrangement of 
sp$^2$--carbon atoms in GPNL creates a two--dimensional 
lattice with regularly spaced sized pores which is notably
larger than the kinetic diameters of H$_2$, CO$_2$, and CH$_4$, 
facilitating the favorable diffusion of these molecules. 
This mechanism highlights the potential of GPNL is a 
promising material for gas separation, particularly 
for CH$_4$, as supported by our MD simulations for CO$_2$ 
purification as well.

\begin{figure}[t!]
    \centering
    \includegraphics[width=0.48\textwidth]{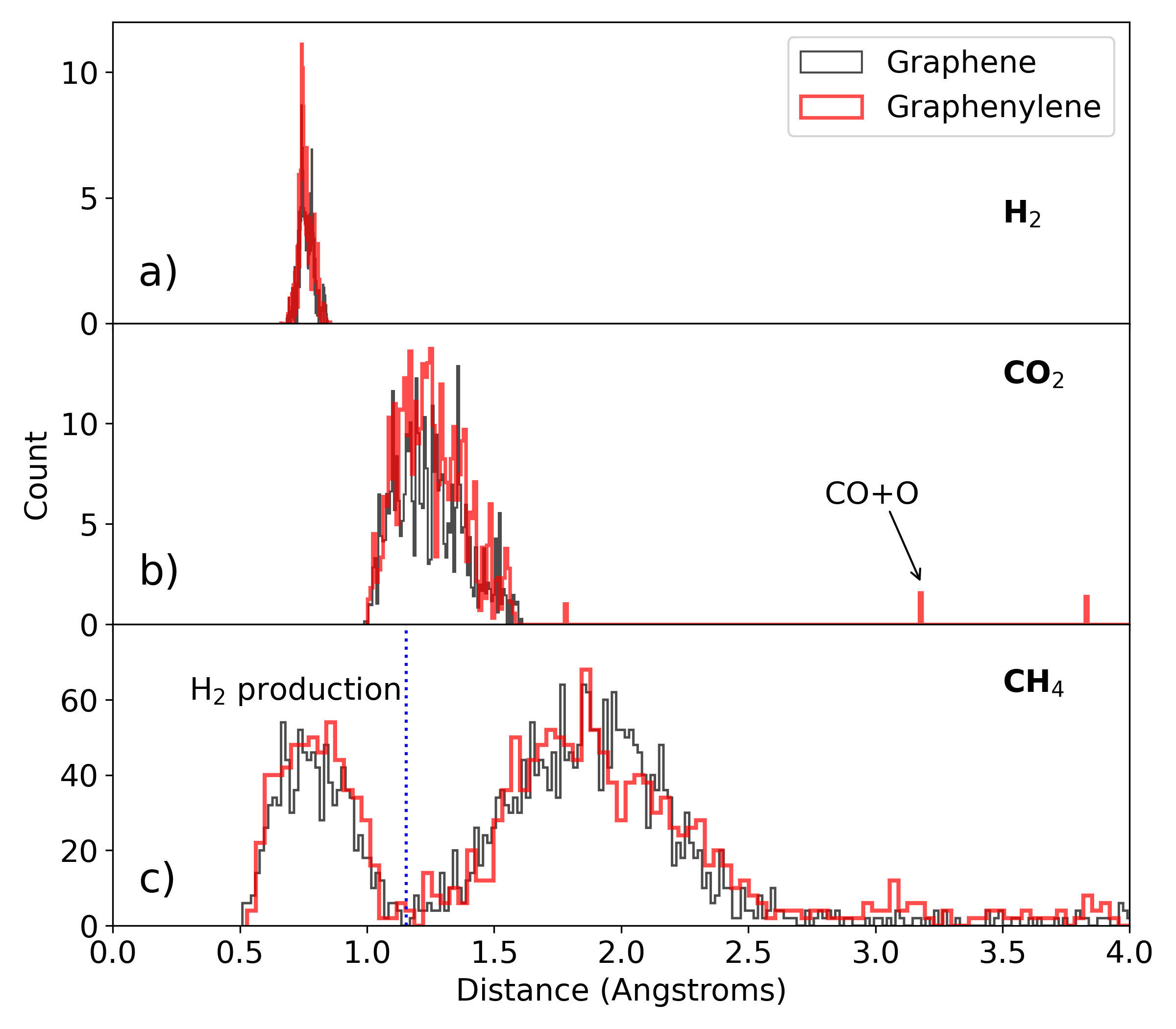}
    \caption{Analysis of reflected molecules from graphene and graphenylene surfaces. Upon collision, H$_2$ molecules are reflected without breaking their bonds, but changing their angular momentum. Most of the CO$_2$ molecules are reflected, while a fraction of them undergo dissociation into CO and O. Interestingly, the collision with the surfaces results in the production of H$_2$ for the majority of CH$_4$ molecules.}
    \label{fig:reflected}
\end{figure}

Figure \ref{fig:reflected} presents the analysis of reflected 
molecules from the surfaces of graphene and GPNL at an 
impact energy of 8 eV. 
The analysis involves determining the internuclear distance
between atoms in the final frame of the simulations.
In the case of H$_2$ molecules (Figure \ref{fig:reflected}a), we 
observe a uniform distribution of the internuclear distance
around the bond length of 0.74 Å. 
This distribution arises 
from the excitation of vibrational and rotational states 
due to the exchange of kinetic energy during a collision
with the surfaces, as previously discussed.
For CO$_2$ molecules, we observe the splitting of the molecules
into CO and O, with a majority exhibiting a homogeneous distribution 
around 
1.43 Å for the internuclear distance which corresponds to the CO 
bond length. Oxygen atoms are identified with 
an internuclear distance larger than 2.5 Å.
Finally, CH$_4$ molecules undergo splitting into CH$_2$ and H$_2$ 
molecules with the highest probabilities, with CH$_3$+H dissociation 
with a low probability. 
Molecular hydrogen is characterized by an internuclear 
distance of around 0.74 Å and a higher degree of excitation in 
vibrational states. 
We identify CH$_2$ (methylene) molecules as the main dissociation
channel with a peak in the histogram at $~1.8$\AA{} being the 
internuclear distance between H atoms.
The porosity of GPNL makes it a more promising 
candidate for H$_2$ production compared to graphene, 
as demonstrated by MD simulations.
The histogram illustrates the distribution of reflected 
molecules with different bond lengths corresponding to 
their excited vibrational states. For instance, when 
H$_2$ molecules are emitted onto the 2D materials, 
they tend to vibrate towards states close to their 
ground state. On the other hand, H$_2$ molecules originating
from the dissociated methane exhibit various bond lengths
due to their excited vibrational and rotational states. 
This concept also applies to CO$_x$ and CH$_x$ molecules.

\subsection{Optical Absorption Spectra, electron transport and sensitivity}
\label{subsec:sensitivity}

Before performing the optical absorption 
calculations, We conducted unconstrained optimizations of the
carbon sheet 
in the presence of various molecules and atoms, each in different 
scenarios. In instances where a physisorption pathway was relevant, 
constrained optimization calculations were performed to establish 
the optimal bond lengths. This approach was adopted to simplify the 
computational process, allowing for a more efficient and accurate 
determination of the atomic configurations.

The optical spectra are calculated
for different cases of molecular adsorption: 
1) For hydrogen molecules, we considered a single H$_2$ 
molecule positioned above a carbon atom, as well as two hydrogen 
atoms positioned above two different carbon atoms, where the 
interaction between them is minimal; the obtained results 
are in a good agreement with experimental and ab--initio data where 
the absorption intensities of the interband transitions occurring
in the Dirac band (mid-IR and visible) 
\cite{LEE2016109,Keith,ZHANG2018137}.
2) For CO$_2$ and its split CO+O molecule, we positioned the CO$_2$ 
molecule above a carbon atom, and for the resulting CO+O system, 
we placed it above two different carbon atoms. 
The effect of carbon dioxide on the graphene sheet is similar
to that observed for the H$_2$ molecule. For GPNL, 
the influence decreases compared to the pristine case. 
However, the split of CO+O enhances the absorbance of the graphene sheet 
at the same wavelength as the CO$_2$ molecule. For the GPNL case, 
the absorbance rate is lower than that observed in the carbon
dioxide case. 
3) For methane molecules, we investigated several 
configurations: CH$_4$, CH$_2$+H$_2$ (molecular hydrogen and 
methylene as a main dissociation channel), CH$_2$+H+H, and CH$_2$,
as they are crucial for sensor development 
\cite{GUI2020113959, zhu2016theoretical}.

\begin{figure}[t!]
    \centering
    \includegraphics[width=0.48\textwidth]{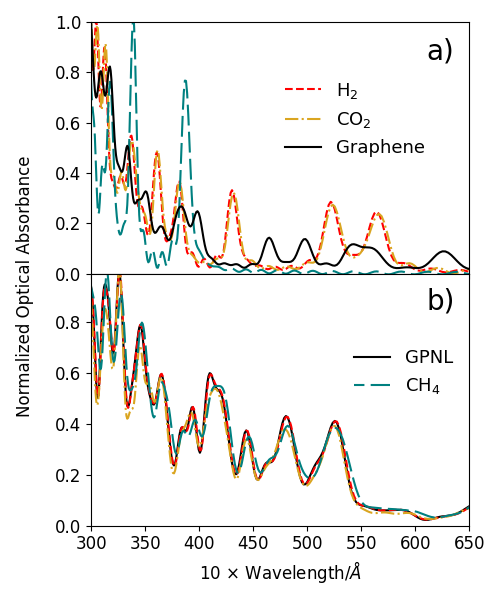}
    \caption{Optical absorbance spectra of graphene (a) and graphenelyne (b) in the visible range, considering H$_2$, CO$_2$, and CH$_4$ molecules adsorbed by the surfaces. The presence of split CH$_4$ molecules significantly impacts both surfaces.}
    \label{fig:optical}
\end{figure}

Figure \ref{fig:optical} presents the normalized optical absorption
spectrum of graphene (a) and GPNL (b) obtained using
the Liouville--von Neumann equation for the hydrogen molecule, 
methane, and carbon dioxide and the rest of the results are 
displayed in the SM. These findings reveal that pristine 
graphene doesn't interact with visible light, as expected
\cite{Hashemi_2013}. 
However, when it's paired with a hydrogen molecule or carbon 
dioxide, graphene becomes more effective at absorbing visible
light. 
In contrast, the addition of a methane molecule primarily
enhances its ultraviolet absorption capabilities. 
Furthermore, when it comes to the GPNL, the presence of attached 
molecules doesn't exert any discernible effect on the optical 
properties of the carbon sheet. The structural characteristics of 
the pores do not significantly enhance light absorption, and its 
band gap doesn't cause excessive scattering of visible light. This 
underscores the material's stability in maintaining its optical 
properties across diverse conditions. 
In the SM, we noticed that among all the configurations, 
the optical absorbance is
maximized in the 300--450 wavelength range for the 
CH$_2$+H+H case, attributed to the bonding between H 
atoms and the C atom of graphene. In contrast, in the 
case of GPNL, CH$_2$ increases the optical 
absorbance in the range of 400 to 550 wavelengths due to the 
system's hybridization and modifying 
the DOS.

Advancements in 2D materials have expanded the scope of potential 
applications, particularly in energy harvesting and storage, due 
to their high electron transport efficiency and extensive surface 
area with numerous active sites \cite{KUMAR2023232256}. In our 
study, we adopted a computational approach based on a $\pi$-
orbital tight-binding Hamiltonian to simulate the electrical 
transport phenomena of graphene and GPNL. We employed the 
SCC-DFTB method along with Non--equilibrium Green's functions 
\cite{Pecchia_2008}.
In the tight--binding representation, the interaction between 
atoms is limited to a finite range. 
We can solve the contact self--energy function, also known 
as the surface Green's function, for 
the matrix block corresponding to the atoms near the extended 
device region using a recursive algorithm \cite{D0CP04188B}. 
This allows us to 
accurately describe the electrical transport properties of the 
materials and analyze their behavior under various conditions.

\begin{figure}[b!]
    \centering
    \includegraphics[width=0.5\textwidth]{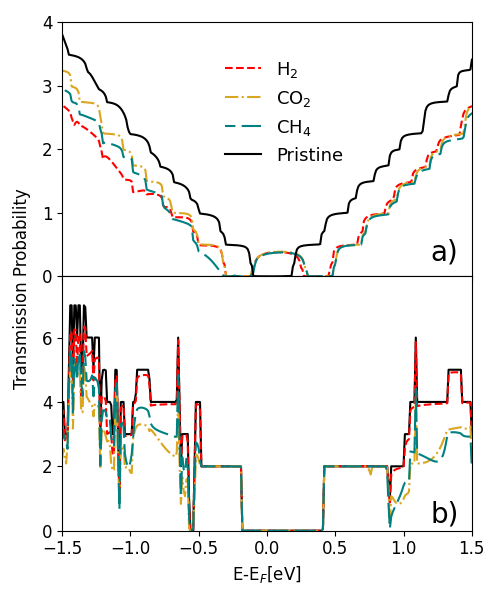}
    \caption{The total transmission probabilities summed over 
    all channels in the nanoribbon graphene direction in a) 
    and through the pores of graphenelyne in b) for the 
    pristine surface and with different molecules 
    observed during the bombardment simulations.}
    \label{fig:transmissionProb}
\end{figure}

Figure \ref{fig:transmissionProb} illustrates our findings 
for graphene in a) and GPNL in b)
that are in good agreement with reported results for 
pristine graphene in the literature and by Villegas 
et al. \cite{D0CP04188B} for the pristine GPNL; 
by adding different molecules it is revealed the preferential 
adsorption of extrinsic chemical 
species like CH$_4$, CO$_2$, and H$_2$ at interdomain sites, 
leading to a significant enhancement of scattering effects in 
both graphene and GPNL. To delve into this intriguing 
behavior, we employed the non--pulsating direct current (NPDC) 
waveform, known for its ability to maintain the stability of 
adsorbates even under ultrahigh--vacuum conditions. Unlike PDC 
waves, which exhibit continuous voltage fluctuations, NPDC waves 
maintain a constant voltage, ensuring the stability of the 
adsorbed species within the scope of our computational approach. 
For graphene, it is noteworthy that the 
adsorption of individual hydrogen, carbon dioxide, and methane 
molecules leads to a reduction in the transmission probability. 
This reduction has a direct impact on the density of states
within the altered system. Additionally, it is observed that 
the effects on the 
DOS of the armchair--wise graphene sheet is associated 
with the bond formed between carbon atoms and these molecules; 
which are reflected in the transmission probability of two symmetric 
minima around the Fermi.
On the other hand, when considering the GPNL sheet, the presence of 
singly bonded molecules to carbon atoms does not appear to induce 
substantial modifications in the transmission probability. 
However, there is a decrease in the transmission probability 
within the energy range of approximately ±1.5 to 1.0 eV. This
observation highlights the potential of GPNL as a promising 
candidate for the development of gas sensors, emphasizing its 
sensitivity to changes in its surrounding gaseous environment.
Since, the strong mechanical properties play a key role in 
maintaining the structural integrity of porous frameworks, 
preventing their shrinkage or collapse. Therefore, the 
presence of channels and pores facilitates rapid electrolyte 
diffusion, leading to an augmentation in electrical conductivity 
as shown by our results.

\begin{figure}[t!]
    \centering
    \includegraphics[width=0.5\textwidth]{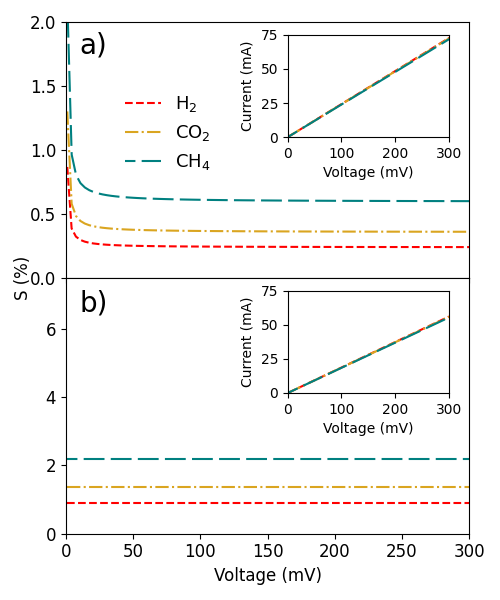}
    \caption{Sensitivity of graphene in a) 
    and graphenelyne in b) considering different molecules 
    and atoms adsorbed. The inset shows the linear current-voltage
    (I-V) characteristics in the range of 0--300 mV.
    }
    \label{fig:I-U}
\end{figure}


Fig \ref{fig:I-U} shows the difference 
of the current for each molecule $X$ as a function of
the voltage as:
\begin{equation}
    S = 100 \% \frac{\big| I_{X}-I_g \big|}{I_{X}},
\end{equation}
with $I_{X}$ of each molecule and the surface and 
$I_g$ the current of the graphene in a) 
and GPNL in b). Noticing that the adsorption of CH$_x$ compounds 
increases the sensitivity of the surfaces.
The tunneling currents for the sheets as a function of 
the voltage for various adsorbed molecules are shown in 
the inset plots by I–-U characteristics graphs for 
voltages below 300 mV.
Our results are in a qualititive good agrement with reported 
experimental data \cite{SHABAN20194510} showing 
a decrease of 25 mV for GPNL and an efficiency increas of 
4\% for methane single molecules. 
Thus, the porous structure of GPNL offers distinct 
advantages, including increased surface area, reduced density, 
and improved accessibility to guest objects. 
This structure is highly suitable for applications involving 
light absorption and electron/ion transport. 
Specifically, it shortens the migration path of charge carriers
from the point of generation to the active surface, 
thereby facilitating electron migration to the surface.
Therefore, we suggest that GPNL based materials can be 
better candidates than graphene ones for use in field effect 
transistors, as Zener diodes, and faster sensors.

\section{Conclusion}
\label{sec:Conclusion}

In this study, we utilize computer simulations to explore the gas 
separation mechanism and its impact on the optical and electronic 
properties of graphene and graphenelyne sheets. Our focus is 
specifically 
on the emission behavior of CH$_4$ and CO$_2$ molecules. To 
conduct these 
simulations, we employ a quantum-classical molecular dynamics 
approach, 
utilizing the SCC--DFTB method with van der Waals corrections. 
These 
corrections are essential to accurately capture the dissociation, 
chemisorption, and molecule formation processes involved in the 
dynamics.

We analyze the probabilities of transmission, reflection, and 
adsorption 
of the emitted molecules. Our results highlight that GPNL 
exhibits significant advantages in gas separation compared to graphene. 
Specifically, we find that GPNL enables efficient separation of 
CO$_2$ into CO+O and CH$_4$ into CH$_2$+H$_2$ with the highest probabilities to be dissociated. The porosity of GPNL enhances gas 
separation rates, facilitates CO$_2$ purification, and promotes hydrogen 
production from methane.
By conducting electron transport calculations with the 
non--equilibrium green function method, we noticed that 
hydrocarbon like CH$_2$, and CH$_4$ 
have the most effects on the electron 
transport mechanisms for both 2D materials. 

\section*{Acknowledgements}
We acknowledge support from the European Union Horizon 2020 
research
and innovation program under grant agreement no. 857470,
from the European Regional Development Fund via 
the Foundation for Polish Science International 
Research Agenda PLUS program grant No. MAB PLUS/ 2018/8. 
We acknowledge the computational resources 
 provided by the High Performance Cluster at the National Centre 
 for Nuclear Research and 
 the Interdisciplinary Centre for Mathematical and
 Computational Modelling (ICM) University of Warsaw under 
 computational allocation no g91--1427.

\bibliography{references}

\providecommand{\newblock}{}
\begin{thebibliography}{10}
\expandafter\ifx\csname url\endcsname\relax
  \def\url#1{{\tt #1}}\fi
\expandafter\ifx\csname urlprefix\endcsname\relax\def\urlprefix{URL }\fi
\providecommand{\eprint}[2][]{\url{#2}}

\bibitem{JANA2022108543}
Jana R, Hajra S, Rajaitha P~M, Mistewicz K and Kim H~J 2022 {\em Journal of
  Environmental Chemical Engineering\/} {\bf 10} 108543 ISSN 2213-3437
  \urlprefix\url{https://www.sciencedirect.com/science/article/pii/S2213343722014166}

\bibitem{balandin2011thermal}
Balandin A~A 2011 {\em Nature materials\/} {\bf 10} 569--581

\bibitem{Wang_2022}
Wang B, Gu Y, Chen L, Ji L, Zhu H and Sun Q 2022 {\em Nanotechnology\/} {\bf
  33} 252001

\bibitem{PhysRevB.77.125416}
Leenaerts O, Partoens B and Peeters F~M 2008 {\em Phys. Rev. B\/} {\bf 77}(12)
  125416 \urlprefix\url{https://link.aps.org/doi/10.1103/PhysRevB.77.125416}

\bibitem{C6CP07654H}
You Y, Deng J, Tan X, Gorjizadeh N, Yoshimura M, Smith S~C, Sahajwalla V and
  Joshi R~K 2017 {\em Phys. Chem. Chem. Phys.\/} {\bf 19}(8) 6051--6056

\bibitem{SHABAN20194510}
Shaban M, Ali S and Rabia M 2019 {\em Journal of Materials Research and
  Technology\/} {\bf 8} 4510--4520 ISSN 2238-7854
  \urlprefix\url{https://www.sciencedirect.com/science/article/pii/S223878541831370X}

\bibitem{ALIGAYEV2022355}
Aligayev A, Raziq F, Jabbarli U, Rzayev N and Qiao L 2022 Chapter 17 -
  morphology and topography of nanotubes {\em Graphene, Nanotubes and Quantum
  Dots-Based Nanotechnology\/} Woodhead Publishing Series in Electronic and
  Optical Materials ed Al-Douri Y (Woodhead Publishing) pp 355--420 ISBN
  978-0-323-85457-3
  \urlprefix\url{https://www.sciencedirect.com/science/article/pii/B9780323854573000190}

\bibitem{shanmugam2022review}
Shanmugam V~K, Mensah R~A, Babu K, Gawusu S, Chanda A, Tu Y, Neisiany R~E,
  Försth M, Sas G and Das O 2022 {\em Part. Part. Syst. Charact.\/} {\bf 39}
  2200031

\bibitem{gao2023tunable}
Gao F, Mench{\'o}n R, Garcia-Lekue A {\em et~al.\/} 2023 {\em Communications
  Physics\/} {\bf 6} 115

\bibitem{hirsch2010era}
Hirsch A 2010 {\em Nature materials\/} {\bf 9} 868--871

\bibitem{karfunkel1992new}
Karfunkel H~R and Dressler T 1992 {\em Journal of the American Chemical
  Society\/} {\bf 114} 2285--2288

\bibitem{baughman1987structure}
Baughman R, Eckhardt H and Kertesz M 1987 {\em The Journal of chemical
  physics\/} {\bf 87} 6687--6699

\bibitem{li2009superhard}
Li Q, Ma Y, Oganov A~R, Wang H, Wang H, Xu Y, Cui T, Mao H~K and Zou G 2009
  {\em Physical review letters\/} {\bf 102} 175506

\bibitem{sheng2011t}
Sheng X~L, Yan Q~B, Ye F, Zheng Q~R and Su G 2011 {\em Physical review
  letters\/} {\bf 106} 155703

\bibitem{C5CP04422G}
Lee G, Yang G, Cho A, Han J~W and Kim J 2016 {\em Phys. Chem. Chem. Phys.\/}
  {\bf 18}(21) 14198--14204

\bibitem{balaban1968chemical}
Balaban A, Rentia C~C and Ciupitu E 1968 {\em Revue Roumaine de Chimie\/} {\bf
  13} 231--+

\bibitem{balaban1994diamond}
Balaban A, Klein D and Folden C 1994 {\em Chemical Physics Letters\/} {\bf 217}
  266--270

\bibitem{C2TC00006G}
Song Q, Wang B, Deng K, Feng X, Wagner M, Gale J~D, MĂĽllen K and Zhi L 2013
  {\em J. Mater. Chem. C\/} {\bf 1}(1) 38--41

\bibitem{D0CP04188B}
Villegas-Lelovsky L and Paupitz R 2020 {\em Phys. Chem. Chem. Phys.\/} {\bf
  22}(48) 28365--28375 \urlprefix\url{http://dx.doi.org/10.1039/D0CP04188B}

\bibitem{du2017new}
Du Q, Tang P, Huang H {\em et~al.\/} 2017 {\em Scientific Reports\/} {\bf 7}
  40796

\bibitem{zhang2019art}
Zhang R and Jiang J 2019 {\em Frontiers in Physics\/} {\bf 14} 13401

\bibitem{brunetto2012nonzero}
Brunetto G, Autreto P, Machado L, Santos B, Dos~Santos R~P and Galvao D~S 2012
  {\em The Journal of Physical Chemistry C\/} {\bf 116} 12810--12813

\bibitem{KOCHAEV2021109999}
Kochaev A, Meftakhutdinov R, Sibatov R and Timkaeva D 2021 {\em Computational
  Materials Science\/} {\bf 186} 109999 ISSN 0927-0256
  \urlprefix\url{https://www.sciencedirect.com/science/article/pii/S0927025620304900}

\bibitem{Martins2022}
Martins N~F, Fabris G~S~L, Albuquerque A~R, Paupitz R and Sambrano J~R 2022
  {\em Graphenylene-Like Structures as a New Class of Multifunctional Materials
  Alternatives to Graphene\/} (Cham: Springer International Publishing) pp
  209--230

\bibitem{D0RA04286B}
Mahdizadeh S~J and Goharshadi E~K 2020 {\em RSC Adv.\/} {\bf 10}(41)
  24255--24264

\bibitem{xu2017inorganic}
Xu J, Zhou S, Sang P, Li J and Zhao L 2017 {\em Journal of Materials Science\/}
  {\bf 52} 10285--10293

\bibitem{zhu2016theoretical}
Zhu L, Jin Y, Xue Q, Li X, Zheng H, Wu T and Ling C 2016 {\em Journal of
  Materials Chemistry A\/} {\bf 4} 15015--15021

\bibitem{rezaee2020graphenylene}
Rezaee P and Naeij H~R 2020 {\em Carbon\/} {\bf 157} 779--787

\bibitem{motallebipour2021graphenylene}
Motallebipour M~S and Karimi-Sabet J 2021 {\em Physical Chemistry Chemical
  Physics\/} {\bf 23} 14706--14715

\bibitem{Sangwan}
Sangwan V~K and Hersam M~C 2018 {\em Annual Review of Physical Chemistry\/}
  {\bf 69} 299--325

\bibitem{D1CP01890F}
Zhang J, Liu L, Yang Y, Huang Q, Li D and Zeng D 2021 {\em Phys. Chem. Chem.
  Phys.\/} {\bf 23}(29) 15420--15439

\bibitem{CHUNG2023215066}
Chung W~T, Mekhemer I~M, Mohamed M~G, Elewa A~M, EL-Mahdy A~F, Chou H~H, Kuo
  S~W and Wu K~C~W 2023 {\em Coordination Chemistry Reviews\/} {\bf 483} 215066
  ISSN 0010-8545
  \urlprefix\url{https://www.sciencedirect.com/science/article/pii/S0010854523000553}

\bibitem{WandOganov}
Wang Z, Zhou X~F, Zhang X, Zhu Q, Dong H, Zhao M and Oganov A~R 2015 {\em Nano
  Letters\/} {\bf 15} 6182--6186

\bibitem{Keith}
Whitener Keith~E J 2018 {\em Journal of Vacuum Science \& Technology A\/} {\bf
  36} 05G401

\bibitem{DFTBplus}
Hourahine B, Aradi B, Blum V, Bonafe F {\em et~al.\/} 2020 {\em The Journal of
  Chemical Physics\/} {\bf 152} 124101

\bibitem{Qiang}
Cui Q, Elstner M, Kaxiras E, Frauenheim T and Karplus M 2001 {\em The Journal
  of Physical Chemistry B\/} {\bf 105} 569--585

\bibitem{Santos_2021}
Santos E and Schmickler W 2021 {\em Journal of Physics: Condensed Matter\/}
  {\bf 33} 504001 \urlprefix\url{https://dx.doi.org/10.1088/1361-648X/ac28c0}

\bibitem{ARUNRAGSA2020107790}
Arunragsa S, Seekaew Y, Pon-On W and Wongchoosuk C 2020 {\em Diamond and
  Related Materials\/} {\bf 105} 107790 ISSN 0925-9635
  \urlprefix\url{https://www.sciencedirect.com/science/article/pii/S0925963520300674}

\bibitem{fabris2018theoretical}
Fabris G, Marana N, Longo E and Sambrano J 2018 {\em Journal of Solid State
  Chemistry\/} {\bf 258} 247--255

\bibitem{martins2021new}
Martins N~F, Fabris G~S, Albuquerque A~R and Sambrano J~R 2021 {\em FlatChem\/}
  {\bf 30} 100286

\bibitem{10.1063/1.1329889}
Elstner M, Hobza P, Frauenheim T, Suhai S and Kaxiras E 2001 {\em The Journal
  of Chemical Physics\/} {\bf 114} 5149--5155

\bibitem{DOMINGUEZGUTIERREZ2018189}
Dominguez-Gutierrez F~J, Krstic P~S, Irle S and Cabrera-Trujillo R 2018 {\em
  Carbon\/} {\bf 134} 189--198

\bibitem{C7TC00976C}
Novotny M, Dominguez-Gutierrez F~J and Krstic P 2017 {\em J. Mater. Chem. C\/}
  {\bf 5}(22) 5426--5433

\bibitem{aluminadftb}
Dominguez-Gutierrez F~J, Aligayev A, Huo W, Chourashiya M, Xu Q and
  Papanikolaou S {\em physica status solidi (b)\/} {\bf n/a} 2200567

\bibitem{C8CP04625E}
Marquez D~M and SĂˇnchez C~G 2018 {\em Phys. Chem. Chem. Phys.\/} {\bf
  20}(41) 26280--26287

\bibitem{B926051J}
Oviedo M~B, Negre C~F~A and SĂˇnchez C~G 2010 {\em Phys. Chem. Chem. Phys.\/}
  {\bf 12}(25) 6706--6711

\bibitem{Pecchia_2008}
Pecchia A, Penazzi G, Salvucci L and Carlo A~D 2008 {\em New Journal of
  Physics\/} {\bf 10} 065022
  \urlprefix\url{https://dx.doi.org/10.1088/1367-2630/10/6/065022}

\bibitem{10.1063/5.0116092}
Lee G, Hong I, Ahn J, Shin H, Benali A and Kwon Y 2022 {\em The Journal of
  Chemical Physics\/} {\bf 157} 144703

\bibitem{LEE2016109}
Lee C, Leconte N, Kim J, Cho D, Lyo I~W and Choi E 2016 {\em Carbon\/} {\bf
  103} 109--114 ISSN 0008-6223
  \urlprefix\url{https://www.sciencedirect.com/science/article/pii/S000862231630197X}

\bibitem{ZHANG2018137}
Zhang W, Lu W~C, Zhang H~X, Ho K and Wang C 2018 {\em Carbon\/} {\bf 131}
  137--141 ISSN 0008-6223
  \urlprefix\url{https://www.sciencedirect.com/science/article/pii/S0008622318301052}

\bibitem{GUI2020113959}
Gui Y, Peng X, Liu K and Ding Z 2020 {\em Physica E: Low-dimensional Systems
  and Nanostructures\/} {\bf 119} 113959 ISSN 1386-9477
  \urlprefix\url{https://www.sciencedirect.com/science/article/pii/S1386947719313116}

\bibitem{Hashemi_2013}
Hashemi M, Farzad M~H, Mortensen N~A and Xiao S 2013 {\em Journal of Optics\/}
  {\bf 15} 055003
  \urlprefix\url{https://dx.doi.org/10.1088/2040-8978/15/5/055003}

\bibitem{KUMAR2023232256}
Kumar M~R, Singh S, Fahmy H~M, Jaiswal N~K, Akin S, Shalan A~E, Lanceros-Mendez
  S and Salado M 2023 {\em Journal of Power Sources\/} {\bf 556} 232256 ISSN
  0378-7753
  \urlprefix\url{https://www.sciencedirect.com/science/article/pii/S0378775322012332}

\end{thebibliography}
\bibliographystyle{iopart-num}

\end{document}